\documentclass{article}
\usepackage{bm,latexsym,amsmath,amsfonts,fancyhdr,color,graphicx,amssymb,axodraw}
\usepackage[a4paper,bottom=2.5cm,top=3.5cm,head=5mm,width=17.5cm,dvipdfm]{geometry}
\usepackage[usenames,dvipsnames,svgnames,table]{xcolor}
\usepackage[ps2pdf,
            colorlinks=true,
            linkcolor=blue,
            urlcolor=red,
            citecolor=green,          
            bookmarks=true,
            bookmarksnumbered=true,
            breaklinks=true,
            pdfpagemode=Fullscreen,
            pdfstartview=FitBH]{hyperref}
\usepackage[dotinlabels]{titletoc}
\usepackage{titlesec}

\newcommand{\titledef}{The Georgi Algorithms of Jet Clustering} % Insert Title here!!!
\newcommand{\vmin}{v^2_{min}}

\numberwithin{equation}{section}
\allowdisplaybreaks[4]

\titlelabel{\thetitle.\quad \hspace{-0.8em}}
\titlecontents{section}
              [1.5em]
              {\vspace{4mm} \large \bf}
              {\contentslabel{1em}}
              {\hspace*{-1em}}
              {\titlerule*[.5pc]{.}\contentspage}
\titlecontents{subsection}
              [3.5em]
              {\vspace{2mm}}
              {\contentslabel{1.8em}}
              {\hspace*{.3em}}
              {\titlerule*[.5pc]{.}\contentspage}
\titlecontents{subsubsection}
              [5.5em]
              {\vspace{2mm}}
              {\contentslabel{2.5em}}
              {\hspace*{.3em}}
              {\titlerule*[.5pc]{.}\contentspage}

\hypersetup{ pdfauthor = {Shao-Feng Ge},
	     pdftitle = {\titledef}, % Insert title here!!!
	     pdfsubject = {}, % Insert subject here!!!
             pdfkeywords = {}, % Insert keywords here!!!
	     pdfcreator = {LaTeX with hyperref package},
	     pdfproducer = {dvips + ps2pdf} }

\definecolor{gesfpurple}{rgb}{0.47,0.19,0.42}

\definecolor{gesflanse}{rgb}{0.00,0.50,0.50}

\definecolor{gesfblue}{rgb}{0.08,0.42,0.76}

\definecolor{gesfred}{rgb}{1,0,0}

\definecolor{gesfwhite}{rgb}{1,1,1}

\definecolor{gesfblack}{rgb}{0,0,0}

\newcommand{\gsec}[1]{{\hypersetup{linkcolor=red}Sec.~\ref{#1}\hypersetup{linkcolor=blue}}}
\newcommand{\gfig}[1]{{\hypersetup{linkcolor=violet}Fig.~\ref{#1}\hypersetup{linkcolor=blue}}}

\begin{document}

\fontsize{11pt}{13pt}\selectfont

%\title{\textbf{\huge \titledef}}
\title{\begin{flushright}
       \mbox{\normalsize KEK-TH-1762}\\[-3mm]
       \mbox{\normalsize arXiv:1408.3823}
       \end{flushright}
       \vskip 20pt
       \textbf{\huge \titledef}} % Insert title here!!!
\author{{\large Shao-Feng Ge} \footnote{\large gesf02@gmail.com} \\[3mm]
        {\large Max-Planck-Institut f\"{u}r Kernphysik, Heidelberg 69117, Germany} \\[2mm]
        {\large KEK Theory Center, Tsukuba 305-0801, Japan}}
\date{\today}

\maketitle

%\thispagestyle{fancy}
%\lhead{Note by Shao-Feng Ge}
%\chead{Created on \cdate}
%\rhead{Modified on \today}
%\lfoot{}
%\cfoot{\thepage}
%\rfoot{}
%
%\pagestyle{fancy}
%\lhead{Note by Shao-Feng Ge}
%\chead{Created on \cdate}
%\rhead{Modified on \today}
%\lfoot{}
%\cfoot{\thepage}
%\rfoot{}

\begin{abstract}
% Insert abstract here!!!
\normalsize
We reveal the direct link between the jet clustering algorithms recently 
proposed by Howard Georgi and parton shower kinematics, providing firm 
foundation from the theoretical side. The kinematics of this class of elegant 
algorithms is explored systematically for partons with arbitrary masses and 
the jet function is generalized to $J^{(n)}_\beta$ with a jet function index 
$n$ in order to achieve more degrees of freedom. Based on three basic requirements that,
the result of jet clustering is process-independent and hence logically consistent, for softer subjets the 
inclusion cone is larger to conform with the fact that parton shower tends 
to emit softer partons at earlier stage with larger opening angle, and 
that the cone size cannot be too large in order
to avoid mixing up neighbor jets, we derive constraints on the jet function
parameter $\beta$ and index $n$ which are closely related to cone size cutoff. 
Finally, we discuss how jet function values can be made invariant under Lorentz boost.
\end{abstract}

%\hypersetup{linkcolor=black}
%\tableofcontents
%\hypersetup{linkcolor=red}

%\vspace{1cm}

\section{Introduction}
\label{sec:intro}

Due to confinement, partons can not be observed directly. The high-energy partons
produced in hard scattering experience shower process first, splitting into low-energy
partons which further fragment into low-energy hadrons. Then, the information of hadrons 
is experimentally measured directly, with the information at the parton level
buried in sprays of hadrons and needs to be reconstructed. Jet is a very 
useful tool for this purpose \cite{Weinberg77} and various algorithms 
have been proposed. The list includes the {\it longitudinally 
invariant $k_t$ algorithm} \cite{Catani93,Ellis93}, the {\it Cambridge/Aachen 
(C/A) algorithm} \cite{Dokshitzer97, Wobisch98, WobischThesis}, the 
{\it anti-$k_t$ algorithm} \cite{Cacciari08}, and the {\it Durham algorithm}
\cite{Catani91}, with different features. For details, please also refer 
to comprehensive reviews \cite{Moretti98, Blazey00, Ellis07, Salam09, Ali10}.
All these jet definitions have a pair-wise feature since the criterion 
whether two jets should be merged into a single one is based on the 
distance between them. Only with two jets, a distance can be defined. The difference
between different algorithms is basically the away of evaluating the distance.

Recently, a new class of algorithms have been proposed \cite{Georgi14} by Howard Georgi. 
In these {\it Georgi algorithms} of jet clustering, 
a {\it jet function} is defined in terms of the jet momentum 
$P_\alpha = (E_\alpha, {\bf P}_\alpha) \equiv \sum_{i \in \alpha} p_i$ as,
\begin{equation}
  J_\beta(P_\alpha)
\equiv
  E_\alpha
- \beta \frac {P^2_\alpha} {E_\alpha}
=
  E_\alpha
\left(
  1
- \beta \frac {P^2_\alpha}{E^2_\alpha}
\right) \,,
\label{eq:Jbeta}
\end{equation}
motivated by the observation that jet emerging leads to fast increase in energy $E_\alpha$ and 
slow increase in the jet mass $P^2_\alpha$, which is small in the first place. When clustering,
the jet function should increase with jets merged, which serves
as the criterion of clustering jet in the Georgi algorithms. Note that $\alpha$ 
is a set of parton indices while $\beta > 1$ is a jet function parameter.
As elaborated in \cite{Georgi14}, $\beta$ is closely related to the jet cone
size cutoff. Its value is of the same place as the distance threshold/cutoff in the traditional
pair-wise jet algorithms. 

Different from distance between jets, jet function can be applied on a single 
jet, providing a new feature of evaluating jet globally. With pair-wise jet distance, 
which can be defined and evaluated locally,
the clustering procedure starts from individual hadrons/subjets, 
merging the closest pair to form a new jet. This merging process iterates until 
the distance between any pair of jets is larger than the threshold/cutoff. Since
it starts from individual hadrons, which can be seen as low-level information, 
the traditional jet algorithms follow a bottom-up approach. On the other hand, the 
Georgi algorithms can be implemented in a global way, starting from dividing the
$4 \pi$ solid angle into fiducial regions \cite{Georgi14}. The jet in each fiducial 
region can be found by looking only at the fiducial region plus some border region.
A subjet should be isolated if without it the jet function 
becomes larger. This is a top-down approach. 

Although not revealed in \cite{Georgi14}, it should be noted that, the jet function 
method can also be carried out locally in a bottom-up approach. 
Starting from evaluating jet function for each individual hadrons, a pair of two 
subjets should be merged into a single jet if the value of jet function
increases when doing so. We can choose the pair with the fastest/slowest increase in
jet function to cluster at each step.

In this paper, we first make connection between the jet function (\ref{eq:Jbeta})
and the parton shower kinematics to establish a theoretical foundation for the
Georgi algorithm in \gsec{sec:ps}.
In \cite{Georgi14}, the kinematic properties of the Georgi algorithms have 
been explored analytically for massless partons. We try to generalize the
results to the massive case in \gsec{sec:massive} and provide further 
generalization of the jet function definition (\ref{eq:Jbeta}) in \gsec{sec:Jn},
to achieve more degrees of freedom. At the end, we briefly discuss how the jet 
function behaves under Lorentz boost in \gsec{sec:boost} and
conclude in \gsec{sec:conclusion}.

\section{Connection with Parton Shower}
\label{sec:ps}

The basic motivation behind the definition (\ref{eq:Jbeta}) of the jet function $J_\beta(P_\alpha)$
is the observation that, jet clustering tends to increase the jet energy $E_\alpha$ 
but the jet mass term $P^2_\alpha/E_\alpha$ does not increase that much. Nevertheless,
this key point is not elaborated in \cite{Georgi14}, by assuming that
``combining a collection of lines into a single jet, hence increasing the jet energy, 
if doing so does not increase the jet mass to much''. We will try to establish
a direct connection between the jet function $J_\beta$ as defined in (\ref{eq:Jbeta}) and 
the kinematics of parton shower, illustrating that the basic motivation of $J_\beta$ has sound 
theoretical foundation. 

For both massless \cite{Catani92} and massive \cite{Gieseke03} 
parton shower schemes, virtuality can be reconstructed iteratively,
\begin{equation}
  P^2_\alpha - m^2_\alpha
=
  \frac {p^2_j - m^2_j}{z}
+ \frac {P^2_{\alpha - j} - m^2_{\alpha - j}} {1 - z}
+ z (1-z) t \,,
\label{eq:virtuality}
\end{equation}
where $P_{\alpha - j} \equiv P_\alpha - p_j$ for $1 \rightarrow 2$ splitting 
$\alpha \rightarrow (\alpha -j) + j$. Note that $\alpha$ and $\alpha - j$
are sets of parton indices while $j$ is the index of a single parton. This formula can 
apply generally, with the massless case restored by setting the parton masses $m_\alpha$, 
$m_j$, and $m_{\alpha - j}$ to zero. For final-state parton shower (FSPS), the 
virtuality of the parent parton can be reconstructed from those of 
the child partons, in a recursive way which is in the same direction as jet clustering. 
This procedure traces back to the partons at the end of the parton shower chains, which are 
physical particles and hence on-shell, $p^2_j = m^2_j$. In this way, the virtualities of all
partons can be reconstructed.

In addition to parton masses, which are known, and virtualities that need to be reconstructed, 
there are two parameters, the evolution scale $t$, which takes the same role as time in 
the decay process of an unstable particle, and the energy fraction $z$, which is an analogy
of energy partitioning between the decay products,
\begin{equation}
  z
\equiv
  \frac {E_j}{E_\alpha}
=
  \frac {E_\alpha - E_{\alpha - j}}{E_\alpha} \,,
\label{eq:z}
\end{equation}
taken away by one of the two child partons
\footnote{For convenience, we have adopted the parton shower notation of energy fraction,
which is different from the original notation $E_j / E_\alpha \equiv r_j$
used in \cite{Georgi14}  where $z$ is used to denote the angle between ${\bf P}_\alpha$ and 
${\bf p}_j$, $\cos \theta$, instead.}. In other words, the parton shower is controlled
by these two parameters, $t$ and $z$. The evolution scale $t$ is
related to the transverse momentum of the child partons,
\begin{equation}
  t
\equiv
  \left. \frac {P^2_\alpha - m^2_\alpha}{z (1-z)} \right|_{P^2_j = m^2_j, P^2_{\alpha - j} = m^2_{\alpha - j}}
=
  \frac 1 {z (1-z)}
\left[
  \frac {m^2_j}{z}
+ \frac {m^2_{\alpha-j}}{1-z}
+ \frac {{\bf p}^2_{\perp j}}{z (1-z)} 
- m^2_\alpha
\right] \,.
\label{eq:t1}
\end{equation}
For FSPS, the evolution scale
$t$ and virtualities are positive.  An immediate conclusion is that the 
parent parton has larger virtuality than the child partons. Consequently,
$P^2_\alpha$ increases when clustering, which is the reverse of parton shower.
Nevertheless, parton shower tends to emit soft partons, $z \rightarrow 0$. 
By clustering the child parton $j$, the relative energy increase from $E_{\alpha - j}$
to $E_\alpha$ is proportional to $z$ as defined in (\ref{eq:z}). On the other hand, the 
relative increase in the second term of (\ref{eq:Jbeta}) is suppressed even more,
by a factor of,
\begin{equation}
  \frac 1 {E_\alpha}
\left[
  \frac {P^2_\alpha - m^2_\alpha}{E_\alpha}
- \frac {P^2_{\alpha - j} - m^2_{\alpha - j}}{E_{\alpha - j}}
\right]
=
  \frac {p^2_j - m^2_j}{z E^2_\alpha}
+ z (1 - z) \frac t {E^2_\alpha} \,.
\label{eq:diff-virtuality}
\end{equation}
Since the parton masses are very small, they can be omitted for convenience,
$(P^2_\alpha - m^2_\alpha)/E_\alpha \approx P^2_\alpha / E_\alpha$.
For a soft emission, the parton with index $j$ tends to be a final-state 
particle, $p^2_j - m^2_j \rightarrow 0$, making the first term vanish.
In addition, the evolution scale $t$ decreases much faster than
energy because of angular ordering \cite{Mueller81, Ermolaev81}, appears as
$t_i < (1-z_{i-1})^2 t_{i-1}$ for FSPS, with the indices assigned according 
to the sequence of splittings in parton shower, the smaller the earlier. 
Note that $t_i$ has very small chance of being close to the 
starting scale $(1-z_{i-1})^2 t_{i-1}$ due to suppression by the so-called 
Sudakov factor $\Delta(t)$, which is an analogy of the exponential decrease,
$e^{- \Gamma t}$, in particle decay. Consequently, the order of the second term in 
(\ref{eq:diff-virtuality}) is lower than $\mathcal O(z)$, and the increase in 
$P^2_\alpha/E_\alpha$ is expected to be smaller than the increase in 
$E_\alpha$. In total, the expression inside the parenthesis of (\ref{eq:Jbeta}) is roughly 
constant. This can be made apparent in the expanded form,
\begin{equation}
  J_\beta (P_\alpha)
=
  E_\alpha
\left[
  (1 - \beta)
+ \beta v^2_\alpha
\right] \,.
\label{eq:Jbeta-expanded}
\end{equation}
as a function of the {\it jet velocity} $v_\alpha \equiv |{\bf P_\alpha}|/E_\alpha$,
which does not change much. For an energetic shower, the
partons are highly relativistic, $v_\alpha \approx 1$. Nevertheless, $v_\alpha$ 
can have slight decrease by clustering since $P^2_\alpha / E^2_\alpha = 1 - v^2_\alpha$ increases as indicated by (\ref{eq:diff-virtuality}). The jet function $J_\beta$ 
increases when reversing the parton shower chain, mainly 
because of the increase in the {\it clustering scale} $E_\alpha$, and hence 
can serve as a natural measure for reconstructing the parton shower history.

In the Georgi algorithms, one extra requirement is that, the jet function 
$J_\beta(P_\alpha)$ is positive, imposing a constraint on the jet velocity 
\cite{Georgi14},
\begin{equation}
  v^2_{min}
\equiv
  1 - \frac 1 \beta
\leq
  v^2 
\leq
  1 \,. 
\label{eq:vAlpha}
\end{equation}
Note that $v^2_{min}$ is just a notation. The only constraint is $v^2_{min}$ should be
smaller than $1$, otherwise, (\ref{eq:vAlpha}) would become meaningless. Equivalently, 
$\beta$ should be positive. Depending on the value of $\beta$, $v^2_{min}$ can 
take any value, even negative values. Nevertheless, the value of $\beta$ should not 
be too large. Otherwise, $1 - 1/\beta \approx 1$, hence $v^2 \approx 1$, 
rendering the allowed range of jet velocity to be
highly suppressed and only almost time-like jets can have a positive
jet function. For $\beta < 1$, the
whole range of jet velocity, $0 \leq v^2 \leq 1$, can be covered by positive jet function.

\section{Clustering with Massive Subjets}
\label{sec:massive}

Jet algorithm is a tool to quantify and visualize the parton shower process. 
Hence, it is helpful to provide a geometrical picture of clustering. To achieve this,
an intuitive choice is cone size. With energy and momentum magnitude
fixed, those subjets contained within a certain cone are all clustered. 
This property has already been implicitly 
incorporated \cite{Georgi14} in the jet definition (\ref{eq:Jbeta}) which
only depends on the angle $\theta$ between the jet 3-momentum ${\bf P}_\alpha$ and
the 3-momentum ${\bf p}_j$ of the subjet,
\begin{equation}
  J_\beta(P_\alpha + p_j)
=
  (E_\alpha + E_j)
\left[
  (1 - \beta)
+ \beta \frac {|{\bf P}_\alpha|^2 + 2 |{\bf P}_\alpha| |{\bf p}_j| \cos \theta + |{\bf p}_j|^2}
							{(E_\alpha + E_j)^2}
\right]\,.
\end{equation}
The jet clustering criterion that the jet function increases,
$J_\beta (P_\alpha + p_j) > J_\beta(P_\alpha)$, constrains
the inclusion cone size, as will be explored in detail below.

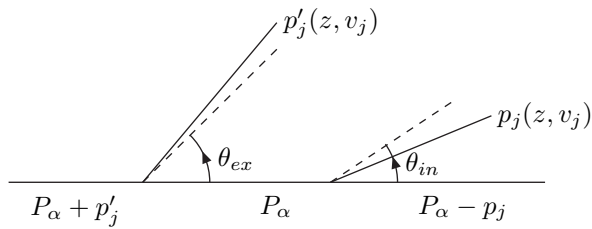
\begin{figure}[h]
\begin{center}
\begin{picture}(200,60)
\Line(0,0)(200,0)
\Text(100,-5)[t]{$P_\alpha$}
\Line(50,0)(100,60)
\Text(103,60)[l]{$p'_j (z, v_j)$}
\DashLine(50,0)(100,50){3}
\ArrowArc(50,0)(25,0,45)
\Text(85,9)[]{$\theta_{ex}$}
\Text(25,-5)[t]{$P_\alpha + p'_j$}
\Line(120,0)(180,25)
\DashLine(120,0)(165,30){3}
\Text(183,25)[l]{$p_j (z, v_j)$}
\ArrowArc(120,0)(25,0,35)
\Text(155,7)[]{$\theta_{in}$}
\Text(170,-5)[t]{$P_\alpha - p_j$}
\end{picture}
\vspace{5mm}
\caption{The inclusion ($\theta_{in}$) and exclusion ($\theta_{ex}$) cones (dashed lines) in
         the sequential clustering of two subjets
         $p_j$ and $p'_j$ (solid lines) with the same energy $z$ and momentum magnitude $v_j$.}
\label{fig:cones}
\end{center}
\end{figure}

Before diving into details, let us first take a look at the big picture
and see what properties we should expect the jet function to have, 
from logical consistency and the property of parton shower, whose 
structure we want to describe by jet clustering. First, the result of jet 
clustering should be independent of the clustering sequence and hence logically 
consistent. For two subjets with the same energy $E_j$,
the same 3-momentum magnitude $|{\bf p}_j|$, and the same angle $\theta$ with
respect to the jet 3-momentum ${\bf P}_\alpha$, both of them should be clustered
if one of them is. Otherwise, one is clustered while the other is not, leaving the 
result process-dependent for a sequential clustering. In other words,
the cone size should not shrink after swallowing a subjet \cite{Georgi14}. What has
not been revealed in \cite{Georgi14} is the fact that this property is also consistent with 
angular ordering \cite{Mueller81, Ermolaev81}, which claims that the 
opening angle between child partons keeps decreasing during parton shower. When 
reversing the process with jet clustering, the inclusion cone size should increase in 
order to accommodate all partons branching from the same chain. Hence, this first 
property is actually a requirement by parton shower, not just by logical consistency.  
It has been depicted in \gfig{fig:cones} as a sequential clustering of two subjets with 
the same energy, parametrized by $z$, and momentum magnitude, denoted as jet velocity
$v_j$. We need to compare the two cones in
two sequential clustering steps, defining the cone at the first as 
inclusion cone, $\theta_{in}$, and the second as exclusion cone, $\theta_{ex}$.
The exclusion cone should not be smaller than the inclusion cone,
$\theta_{ex} \geq \theta_{in}$.
The second property comes from the tendency of parton shower to emit 
softer parton at earlier stage with larger opening
angle \cite{Catani92,Gieseke03}. Consequently, to make jet clustering
approach the real parton shower process, it is necessary to have a larger 
cone size for softer subjet. Although it has been noticed in \cite{Georgi14} that 
``the bound on jet `size' in the sense
of the largest possible angle of a particle in the jet from the jet direction
is determined by the soft particles in the jet'', it is only from parton shower
that we realize this is a ``must''.
We will show that these first two properties can be parameterized with a same 
quantity. Together, they eliminate the parameter region, $\beta < 0$.
The third property is that the inclusion cone cannot be too large. For the 
simplest case of $e^+ e^- \rightarrow jj$ at LEP, the inclusion cone should 
not be larger than half sphere. Otherwise, the two jets cannot be
separated. We will show that this gives a more stringent limit, $\beta > 1$.

In the pioneering work \cite{Georgi14}, the author claims that ``in practice we will typically 
be interested in masses $\sqrt{p^\mu_j p_{j \mu}}$ that are small compared to their 
energies and can be ignored in leading order''.  As pointed out therein,
``this is not necessary for the construction, but it leads to considerable 
simplification''. We will show that it needs not to ignore the parton mass $\sqrt{p^\mu_j p_{j \mu}}$.
In this section, we derive the most general form of the Georgi 
algorithms \cite{Georgi14} by keeping the subjet velocity 
$v_j \equiv |{\bf p}_j|/E_j$ without simplification. 
The massless case can be restored by when $v_j$ approaches $1$.

\subsection{The Inclusion Cone}

Let us start with the clustering of the first subjet with momentum $p_j$
of \gfig{fig:cones} to determine the inclusion cone size in terms of
kinematic variables and possible constraints. For 
convenience of comparison with the clustering of the second subjet $p'_j$,
which will be explored in \gsec{sec:exclusion}, we 
parameterize the clustering criterion on a common ground $P_\alpha$ where $p_j$ 
has already been clustered, but $p'_j$ has not. In other words, $p_j$
is a part of $P_\alpha$, to be exact $j \in \alpha$ where $\alpha$ is a set of 
subjet indices, and the jet momentum before clustering is 
$P_\alpha - p_j$. Jet clustering criterion requires that the jet function 
(\ref{eq:Jbeta}) increases,
\begin{equation}
  J_\beta(P_\alpha)
>
  \max
\left\{
  J_\beta(P_\alpha - p_j),
  J_\beta(p_j)
\right\} \,.
\label{eq:Jcriterion0}
\end{equation}
Note that the jet function increases with respect to both subjets, because
in reality it is impossible to distinguish the two sources.
Using the expanded form (\ref{eq:Jbeta-expanded}) of the jet function, 
these two constraints can be expressed as,
\begin{subequations}
\begin{eqnarray}
  (1 - \beta) + \beta v^2_\alpha
& > &
  (1 - z) \left[ (1 - \beta) + \frac \beta {(1-z)^2} (v^2_\alpha + z^2 v^2_j - 2 z \cos \theta v_\alpha v_j) \right] \,,
\label{eq:criterion0a}
\\
  (1 - \beta) + \beta v^2_\alpha
& > &
  z \left[ (1 - \beta) + \beta v^2_j \right] \,.
\label{eq:criterion0b}
\end{eqnarray}
\label{eq:criterion0}
\end{subequations}
We can see that the second inequality (\ref{eq:criterion0b}) gives a limit on 
the clustered jet velocity,
\begin{equation}
  v^2_\alpha - \vmin
>
  z \left( v^2_j - \vmin \right) \,.
\label{eq:vboundary0}
\end{equation}
This simply indicates that if $J_\beta(p_j)$ is positive,
$J_\beta(P_\alpha)$ is also positive. The jet velocity range is enlarged after
clustering for $\beta > 1$, $v^2_\alpha$ is even closer to $v^2_{min}$ than $v^2_j$, 
due to the $z$ factor in (\ref{eq:vboundary0}) 
which originates from the enhancement contributed by the clustering scale 
$E_\alpha$ in the jet definition (\ref{eq:Jbeta}).

On the other hand, (\ref{eq:criterion0a}) limits the jet cone size,
\begin{equation}
  \cos \theta
>
  \cos \theta_{in}
\equiv
  \frac {\left[ (1-z) \vmin + z v^2_j \right] + v^2_\alpha}
        {2 v_\alpha v_j}
\geq
  \sqrt{ (1-z) \frac {\vmin}{v^2_j} + z } \,.
\label{eq:cone-inclusion}
\end{equation}
Note that the second inequality comes from minimizing $\cos \theta_{in}$ as a 
function of $v_\alpha$, and the equality happens on the boundary 
(\ref{eq:vboundary0}) of $v_\alpha$ if $\beta > 1$. 
For massive subjet, $v^2_j < 1$, the maximal inclusion cone
is smaller than the massless limit.
The most interesting feature is the dependence
on the energy fraction $z$. We can decompose the cone size $\cos \theta_{in}$
(\ref{eq:cone-inclusion}) as a series of $z$,
\begin{equation}
  \cos \theta_{in}
\equiv
  \frac 1 {2 v_\alpha v_j}
\left[
\left(
  v^2_\alpha + \vmin
\right)
+
\left(
  v^2_j
- \vmin
\right) z
\right] \,,
\label{eq:cone-inclusion-z}
\end{equation}
which reduces to the Eq(15) of \cite{Georgi14} in the massless limit, 
$v_j \rightarrow 1$, replacing $v^2_{min}$ with $1 - 1/\beta$, while changing 
the notations, $z \rightarrow r_j$ and $\cos \theta_{in} \rightarrow z$. This 
indicates that the cone increases with decreasing $z$. In other words, the cone 
is larger for softer subjet if (\ref{eq:vAlpha}) is satisfied, which is
exactly what required by parton shower. It provides
a strong support for the requirement on the positiveness of the jet function.

The inclusion region (\ref{eq:cone-inclusion-z}) can also be expressed 
in terms of $\sin (\theta_{in}/2)$ which increases with the cone size,
\begin{equation}
  2 \sin^2 \left( \frac 1 2 \theta_{in} \right)
=
  \frac 1 {2 v_\alpha v_j}
\left[
  (1 - z) \left( v^2_j - \vmin \right)
- (v_\alpha - v_j)^2
\right]
>
  \frac {v_j - v_\alpha} {v_j} \,,
\label{eq:J1-inclusion-bound}
\end{equation}
where the inequality comes from (\ref{eq:vboundary0}). Since virtuality
increases when reversing the parton shower history according to 
(\ref{eq:virtuality}), the velocity decreases, $v_j > v_\alpha$. This
indicates that the inclusion cone cannot be too small.
 If we take the soft and massless limits,
$z \rightarrow 0$ and $v_j, v_\alpha \rightarrow 1$ respectively, the inclusion cone
becomes $\cos \theta_{in} = (1 + v^2_{min})/2$, which together with (\ref{eq:vAlpha})
reduces to the Eq.(17) of \cite{Georgi14}, $\theta_{in} = 2 \arcsin (1/2\sqrt \beta)$.
On the other hand, in the soft limit, $z \rightarrow 0$, massless limit for
the subjet, $v_j \rightarrow 1$, and the lower limit (\ref{eq:vAlpha}) on the
jet velocity after clustering, $v^2_\alpha = v^2_{min} = 1 - 1/\beta$, it reduces to
$\cos \theta_{in} = v_{min} = v_\alpha$ and equivalently,
$ \theta_{in}
=
  2 \arcsin \sqrt{ ( 1 - \sqrt{1 - 1/\beta} ) / 2 }
$, which is the Eq.(18) of \cite{Georgi14}.

\subsection{The Exclusion Cone}
\label{sec:exclusion}

After including the first subjet $p_j$ with certain energy ($z$) and 
3-momentum magnitude ($v_j$), the jet momentum changes from $P_\alpha - p_j$ 
to $P_\alpha$, as shown in \gfig{fig:cones}. This leads to modification of 
the cone size which we will try to derive here. The result is compared with
the one of inclusion cone.
If the jet clustering is self-consistent, a second subjet $p'_j$
with the same energy $z$ and 3-momentum magnitude $v_j$ should also be clustered if
inside the inclusion region (\ref{eq:cone-inclusion}). 
In addition, the largest cone size is established in terms of
the jet parameter $\beta$.

Suppose this second subjet $p'_j$ can not be clustered, namely, the jet 
function (\ref{eq:Jbeta}) decreases if so,
\begin{equation}
  J_\beta(P_\alpha)
>
  \max
\left\{
  J_\beta(P_\alpha + p'_j),
  J_\beta(p'_j)
\right\} \,.
\label{eq:criterion-exclusion}
\end{equation}
From these two constraints we can derive the exclusion cone, parametrized 
with $\theta_{ex}$ as shown if \gfig{fig:cones}. Using the expanded form 
(\ref{eq:Jbeta-expanded}) of the jet function, we can get two inequalities,
\begin{subequations}
\begin{eqnarray}
  (1 - \beta) + \beta v^2_\alpha
& > &
  (1 + z) \left[ (1 - \beta) + \frac \beta {(1+z)^2} (v^2_\alpha + z^2 v^2_j + 2 z \cos \theta' v_\alpha v_j) \right] \,,
\label{eq:criterion-exclusion-a}
\\
  (1 - \beta) + \beta v^2_\alpha
& > &
  z \left[ (1 - \beta) + \beta v^2_j \right] \,.
\label{eq:criterion-exclusion-b}
\end{eqnarray}
\label{eq:criterion-exclusion}
\end{subequations}
Note that (\ref{eq:criterion-exclusion-b}) is exactly (\ref{eq:criterion0b}),
leading to the same constraint (\ref{eq:vboundary0}) on $v^2_\alpha$. But the
cone boundary,
\begin{equation}
  \cos \theta'
<
  \cos \theta_{ex}
\equiv
  \frac {\left[ (1+z) \vmin - z \beta v^2_j \right] + v^2_\alpha}
        {2 v_\alpha v_j} \,,
\label{eq:cone-exclusion}
\end{equation}
is different from the inclusion cone boundary $\cos \theta_{in}$ in 
(\ref{eq:cone-inclusion}). This difference can be traced back to the different signs 
of $p_j$ and $p'_j$ in
the jet functions $J_\beta(P_\alpha - p_j)$ and $J_\beta(P_\alpha + p'_j)$, 
respectively, leading to an effective replacement, $z \rightarrow -z$. Note that
the inclusion cone (\ref{eq:cone-inclusion}) and the exclusion cone
(\ref{eq:cone-exclusion}) are well separated due to the lower limit
(\ref{eq:vAlpha}) on $v_j$,
\begin{equation}
  \cos \theta_{in} - \cos \theta_{ex}
=
  \frac z {v_\alpha v_j}
\left(
  v^2_j - \vmin
\right)
> 
  0 \,.
\label{eq:diff}
\end{equation}
The subjet within the inclusion cone (\ref{eq:cone-inclusion}) with the same energy 
$z$ and 3-momentum magnitude $v_j$ can be readily clustered. Imposing this property 
eliminates the possibility of $v^2_{min} > 1$, or equivalently $\beta < 0$, 
since the jet velocity is bounded by the speed of light from above, $v^2_j \leq 1$.

It should be emphasized that the only difference between the inclusion cone
(\ref{eq:cone-inclusion}) and the exclusion cone (\ref{eq:cone-exclusion}) 
is a sign difference associated with $z$. For soft jet, $z \rightarrow 0$, the 
difference between the two cones actually also characterizes the ability of 
accommodating softer subjet. This can be explicitly seen by comparing the 
expression of $\cos \theta_{in} - \cos \theta_{ex}$ in
(\ref{eq:diff}) and the linear term of $z$ in (\ref{eq:cone-inclusion-z})
whose coefficients differ by only a factor of $2$.
The inclusion region should expand during the jet clustering
process in order to accommodate softer subjet while it is the opposite for
the exclusion region, approaching each other. This is consistent with the
observation in \cite{Georgi14} that, ``the particles not in the jet can only
approach the jet boundary as $r_j \rightarrow 0$''.

Similar to (\ref{eq:J1-inclusion-bound}), the exclusion cone size is bounded by,
\begin{equation}
  2 \sin^2 \left( \frac 1 2 \theta_{ex} \right)
=
  \frac 1 {2 v_\alpha v_j}
\left[
  (1 + z) \left( v^2_j - \vmin \right)
- (v_\alpha - v_j)^2
\right]
<
  1
- \frac {\vmin} {v_\alpha v_j} \,,
\label{eq:J1-exclusion-bound}
\end{equation}
where the inequality comes from (\ref{eq:vboundary0}).
A direct consequence is,
\begin{equation}
  2 \sin^2 \left( \frac 1 2 \theta_{ex} \right)
<
  1 - \vmin
=
  \frac 1 \beta \,,
\label{eq:J1-sin-exclusion}
\end{equation}
since $v_\alpha, v_j \leq 1$. For $\beta \geq 1$, the exclusion cone is smaller than half sphere, 
$\theta_{ex} < 90^\circ$. Take a two-jet event in the center-of-mass 
frame as illustration, for example $e^+ e^- \rightarrow jj$ at LEP, jet-clustering
should reconstruct two jets that are back-to-back. On the other hand,
the two jets can be mixed with each other if $\beta < 1$, rendering the cone
to be larger than half sphere, which is a not good choice. 
This observation can serve as a guide for choosing a reasonable 
value for $\beta$. To recognize an event with more primary jets, $\beta$ should be larger. 
Note that this limit is independent of $z$.

\section{Generalized Jet Function}
\label{sec:Jn}

In the previous section, we have shown that the jet parameter $\beta$ is closely related
cone size cutoff. With larger $\beta$, the cone size becomes smaller. It would be a good
practice to find an extension, providing more degrees of freedom when choosing the cone
size cutoff. Here, we try to develop a possible generalization by introducing {\it jet function
index} $n$, which is also related to cone size as we will elaborate in the remaining part of
this section.

Technically speaking, the generalization comes from the observation that $v^2_\alpha$ appears 
on both sides of (\ref{eq:criterion0a}) and (\ref{eq:criterion-exclusion-a}), where it is 
possible to make a complete cancellation of the $v^2_\alpha$ terms if the prefactor $1-z$ 
is replaced by $(1-z)^2$. The same trick
can be used to remove the factor $1-\beta$ in (\ref{eq:criterion0}) and
(\ref{eq:criterion-exclusion}) by eliminating the prefactor $1-z$. Nevertheless,
the first observation can become true but the latter is not realistic as will
be shown in detail below.

Since the power of the $1-z$ prefactor can be traced back to the power of the
clustering scale $E_\alpha$, to achieve the small tricks we need to generalize 
the jet function (\ref{eq:Jbeta}) as follows,
\begin{equation}
  J^{(n)}_\beta (P_\alpha)
\equiv
  E^n_\alpha
\left(
  1
- \beta \frac {P^2_\alpha}{E^2_\alpha}
\right)
=
  E^n_\alpha
\left[
  (1-\beta)
+ \beta v^2_\alpha
\right] \,,
\label{eq:Jn}
\end{equation}
with an extra jet index $n$.
Accordingly, the jet function (\ref{eq:Jbeta}) can be renamed as
$J_\beta(P_\alpha) \equiv J^{(1)}_\beta(P_\alpha)$. 
For generality, we keep the jet function index $n$ as a free parameter 
in the following derivations. Note that $n$ needs not to be an integer and 
can serve as a jet function parameter as $\beta$. Its value is constrained by
kinematics. As we have argued that the jet function increases mainly
because of the increase in the prefactor $E^n_\alpha$. The jet function index
$n$ cannot be arbitrarily small for the jet function to increase fast enough.
At least, $n$ has to be positive. We will show further constraints in the 
following analysis.

With this generalized jet definition, the limit (\ref{eq:vAlpha}) on jet velocity 
from the requirement that the jet function has to be positive
is still the same. As expected, the $1-z$ and $z$ prefactors in the inclusion 
(\ref{eq:criterion0}) and exclusion (\ref{eq:criterion-exclusion}) criteria 
receives a nontrivial power $n$,
\begin{subequations}
\begin{eqnarray}
  (1 - \beta) + \beta v^2_\alpha
& > &
  (1 \mp z)^n \left[ (1 - \beta) + \frac \beta {(1 \mp z)^2} (v^2_\alpha + z^2 v^2_j \mp 2 z \cos \theta v_\alpha v_j) \right] \,,
\label{eq:nth-criterion-inclusion-a}
\\
  (1 - \beta) + \beta v^2_\alpha
& > &
  z^n \left[ (1 - \beta) + \beta v^2_j \right] \,,
\label{eq:nth-criterion-inclusion-b}
\end{eqnarray}
\end{subequations}
with the sign $\mp$ corresponding to inclusion and exclusion cones, respectively.
From the second inequality, we can get a generalized form of 
the jet velocity constraint (\ref{eq:vboundary0}),
\begin{equation}
  v^2_\alpha - \vmin
>
  z^n 
\left( v^2_j - \vmin \right) \,.
\label{eq:nth-vAlpha-criterion}
\end{equation}
Similarly, $v^2_\alpha$ is contained within the positive jet function region 
(\ref{eq:vAlpha}) if $v^2_j$ already satisfies it. The jet velocity 
range of the $\alpha$-set becomes larger than that of the subjet, due to suppression
from the prefactor $z^n$. 
In soft jet approximation, $z \rightarrow 0$, the difference
can be significant.

The inclusion cone (\ref{eq:cone-inclusion}) and the exclusion cone
(\ref{eq:cone-exclusion}) are constrained by the first inequality
(\ref{eq:nth-criterion-inclusion-a}),
%\begin{subequations}
%\begin{eqnarray}
%  \cos \theta^{(n)}_{in}
%\equiv
%  \frac {[1 - (1-z)^n] \left( 1 - \frac 1 \beta \right) + (1-z)^{n-2} z^2 v^2_j + [(1-z)^{n-2} - 1] v^2_\alpha}
%				{2 z (1-z)^{n-2} v_\alpha v_j} \,,
%\label{eq:nth-inclusion-cone}
%\\
%  \cos \theta^{(n)}_{ex}
%\equiv
%  \frac {[(1+z)^n - 1] \left( 1 - \frac 1 \beta \right) - (1+z)^{n-2} z^2 v^2_j + [1 - (1+z)^{n-2}] v^2_\alpha}
%				{2 z (1+z)^{n-2} v_\alpha v_j} \,.
%\label{eq:nth-exclusion-cone}
%\end{eqnarray}
%\label{eq:nth-cones}
%\end{subequations}
\begin{subequations}
\begin{eqnarray}
  \cos \theta^{(n)}_{in}
\equiv
  \frac 1 {2 z v_\alpha v_j}
\left\{
  \left[ 1 - (1-z)^{2-n} \right]
  \left( v^2_\alpha - \vmin \right)
+
  z^2 \left( v^2_j - \vmin \right)
+ 2 z \vmin
\right\} \,,
\label{eq:nth-inclusion-cone}
\\
  \cos \theta^{(n)}_{ex}
\equiv
  \frac 1 {2 z v_\alpha v_j}
\left\{
  \left[ (1+z)^{2-n} - 1 \right]
  \left( v^2_\alpha - \vmin \right)
-
  z^2 \left( v^2_j - \vmin \right)
+
  2 z \vmin
\right\} \,.
\label{eq:nth-exclusion-cone}
\end{eqnarray}
\label{eq:nth-cones}
\end{subequations}
Then, we can explore the difference between them,
\begin{subequations}
\begin{eqnarray}
  \cos \theta^{(n)}_{in}
- \cos \theta^{(n)}_{ex}
& = &
 \frac
{
  2
- (1 - z)^{2-n}
- (1 + z)^{2-n}
} {2 z v_\alpha v_j}
\left(
  v^2_\alpha - \vmin
\right)
+
  \frac z {v_\alpha v_j}
\left(
  v^2_j - \vmin
\right)
\label{eq:nth-diff-a}
\\
& \geq &
  \frac 1 {2 v_\alpha v_j}
\left\{
  z^{n-1}
\left[
  2
- (1 - z)^{2-n}
- (1 + z)^{2-n}
\right]
+ 2 z
\right\}
\left(
  v^2_j - \vmin
\right) \,.
\label{eq:nth-diff-b}
\end{eqnarray}
\label{eq:nth-diff}
\end{subequations}
If the clustering algorithm is self-consistent, the inclusion 
cone expands after clustering a subjet with the same energy and 3-momentum 
magnitude with the only difference in opening angle. This property makes itself
explicit as the inequality in (\ref{eq:nth-diff}), which is satisfied for 
$n \leq 2$. The self-consistency requirement of jet clustering 
provides an upper limit on the jet function index $n$. Note that, for both 
$n = 1$ and $n = 2$, (\ref{eq:nth-diff-b}) reduces to (\ref{eq:diff}). 

Now let us take a look at the soft region. If the subjet is soft, 
$z \rightarrow 0$, the inclusion and exclusion cones (\ref{eq:nth-cones}) can 
be approximated by an expansion up to the linear order of $z$,
\begin{subequations}
\begin{eqnarray}
  \cos \theta^{(n)}_{in}
\approx
  \frac 1 {2 v_\alpha v_j}
\left\{
  (2-n) \left( 1 - \frac {1-n} 2 z \right)
  \left( v^2_\alpha - \vmin \right)
+
  z \left( v^2_j - \vmin \right)
+
  2 \vmin
\right\} \,,
\label{eq:nth-inclusion-cone-expanded}
\\
  \cos \theta^{(n)}_{ex}
\approx
  \frac 1 {2 v_\alpha v_j}
\left\{
  (2-n) \left( 1 + \frac {1-n} 2 z \right)
  \left( v^2_\alpha - \vmin \right)
-
  z \left( v^2_j - \vmin \right)
+
  2 \vmin
\right\} \,.
\label{eq:nth-exclusion-cone-expanded}
\end{eqnarray}
\label{eq:nth-cones-expanded}
\end{subequations}
The difference (\ref{eq:nth-diff}) between the inclusion and exclusion cones 
is roughly,
\begin{eqnarray}
  \cos \theta^{(n)}_{in}
- \cos \theta^{(n)}_{ex}
\approx
- \frac 1 {2 v_\alpha v_j}
	(2-n)(1-n) z
\left(
  v^2_\alpha - \vmin
\right)
+
  \frac z {v_\alpha v_j}
\left(
  v^2_j - \vmin
\right) \,,
\end{eqnarray}
which is highly suppressed. Nevertheless, overlapping can still happen.
To avoid this tiny chance, the following relation between $v^2_\alpha$
and $v^2_j$ has to be satisfied,
\begin{equation}
  v^2_j - \vmin
\geq
  \frac {(2-n)(1-n)} 2
\left(
  v^2_\alpha - \vmin
\right) \,.
\end{equation}
Together with (\ref{eq:nth-vAlpha-criterion}), we can get,
\begin{equation}
\left[
  \frac 1 {z^n}
- \frac {(2-n)(1-n)} 2
\right]
\left(
  v^2_\alpha - \vmin
\right)
\geq 0 \,,
\end{equation}
which is always true for $n > 0$. The jet-clustering self-consistency
in the soft region also imposes a lower limit on the jet function index $n$.
Since self-consistency in the soft region is directly related to the requirement
that soft emission has a larger inclusion cone, in order to make the jet 
algorithm approach the parton shower evolution, this lower limit can also
be treated as a requirement of the second property.

To see the boundary on the exclusion cone, we need to first check the sign
of $z$ in the expanded form (\ref{eq:nth-exclusion-cone-expanded}),
\begin{eqnarray}
  \cos \theta^{(n)}_{ex}
\approx
  \frac 1 {2 v_\alpha v_j}
\left\{
  (2 - n) \left( v^2_\alpha - \vmin \right)
+ 2 \vmin
+
\left[
  \frac {(2-n)(1-n)} 2 \left( v^2_\alpha - \vmin \right)
- \left( v^2_j - \vmin \right)
\right] z
\right\} .
\label{eq:nth-exclusion-cone-zseries}
\end{eqnarray}
We can see that, by replacing $z$ with the help of (\ref{eq:nth-vAlpha-criterion}) 
the exclusion cone can have a bound like (\ref{eq:J1-sin-exclusion}), which is independent 
of $z$, as long as the coefficient of $z$ in (\ref{eq:nth-exclusion-cone-zseries}) 
is negative. This can be guaranteed for $1 \leq n \leq 2$, 
\begin{eqnarray}
  \cos \theta^{(n)}_{ex}
& \gtrsim &
  \frac 1 {2 v_\alpha v_j}
\left[
  (2 - n) \left( v^2_\alpha - \vmin \right) + 2 \vmin
\right]
\nonumber
\\
& + &
  \frac 1 {2 v_\alpha v_j}
\left[
  \frac {(2-n)(1-n)} 2 \left( v^2_\alpha - \vmin \right)
- \left( v^2_j - \vmin \right)
\right]
\left(
  \frac {v^2_\alpha - \vmin} 
				{v^2_j - \vmin}
\right)^{\frac 1 n} \,.
\label{eq:cos-ex-limit}
\end{eqnarray}
This expression is a little bit too complicated, and we would try to make simplifications. 
Since the partons are quite relativistic, $v_\alpha \approx v_j \approx 1$, (\ref{eq:cos-ex-limit}) 
reduces to,
\begin{equation}
  2 \sin^2 \left( \frac 1 2 \theta^{(n)}_{ex} \right)
\lesssim
  \frac {n (5 - n)} 4 \frac 1 \beta \,.
\end{equation}
This simplification has another advantage of expressing the boundary in terms of
the jet parameter $\beta$ and index $n$.
The result (\ref{eq:J1-sin-exclusion}) can be reproduced with $n=1$. We can see
that both $\beta$ and $n$ are directly related to the kinematic boundary. For the 
cone size to be larger than half sphere, $\beta$ has a lower limit,
\begin{equation}
  \beta
>
  \frac 4 {n (5 - n)}
\geq
  \frac 2 3 \,.
\end{equation}
In the range
$1 \leq n \leq 2$, the coefficient $n(5-n)$ decreases with $n$. Consequently,
$\beta$ should increase with $n$. For the original scheme, $n = 1$, 
$\beta > 1$, leading to $0 < \vmin < 1$. Only part of the jet velocity range can be 
covered which is especially true with more than $2$ jets and $\beta$ further
enhanced. By generalization, $n > 1$, the parameter
$\beta$ can be smaller than $1$, leading to a negative $\vmin$ which can cover
the whole jet velocity range.

Similarly, there is a lower limit on the inclusion cone size,
\begin{eqnarray}
  \cos \theta^{(n)}_{in}
& \lesssim &
  \frac 1 {2 v_\alpha v_j}
\left[
  (2 - n) \left( v^2_\alpha - \vmin \right)
+ 2 \vmin
\right]
\nonumber
\\
& - &
  \frac 1 {2 v_\alpha v_j}
\left[
  \frac {(2-n)(1-n)} 2 \left( v^2_\alpha - \vmin \right)
- \left( v^2_j - \vmin \right)
\right]
\left(
\frac {v^2_\alpha - \vmin}
      {v^2_j - \vmin}
\right)^{\frac 1 n} \,,
\end{eqnarray}
which reduces to,
\begin{equation}
  2 \sin^2 \left( \frac 1 2 \theta^{(n)}_{in} \right)
\gtrsim
  \frac {n (n-1)} 4 \frac 1 \beta \,,
\end{equation}
in the relativistic limit.
For $n \geq 1$, the inclusion cone cannot be arbitrarily small.

\section{Lorentz Boost Invariance}
\label{sec:boost}

From the constraint on the jet velocity (\ref{eq:vAlpha}), we can see that 
$\beta = 1 / (1 - v^2_{min})$ is actually the square of the corresponding boost 
factor $\gamma_{min} = 1 / \sqrt{1 - v^2_{min}}$. This indicates that $\beta$
has close relation with Lorentz boost. It is important to check how Lorentz
boost affects the jet algorithm, especially for highly boosted jets at hadron 
collider like LHC.

The Lorentz boost can be represented by boost factor $\gamma_B$ and the direction
of boosting. At each step of clustering, the jet momentum changes. Not just its 
magnitude is different, but also the direction. There is no uniform transformation 
on the jet momenta. For jet function under consideration, the jet mass, $P^2_\alpha$, 
is invariant, but the jet energy, $E_\alpha$, changes. The rate of energy
change is different from jet to jet. This can be parametrized as,
\begin{equation}
  P^2_\alpha \rightarrow P^2_\alpha \,,
\qquad
  E_\alpha \rightarrow \gamma_\alpha E_\alpha \,.
\end{equation}
The rescaling factor $\gamma_\alpha$ is not universal, but varies from jet to jet 
as a function of the boost factor, $\gamma_B$, and the jet velocity, $v_\alpha$,
\begin{equation}
  \gamma_\alpha
\equiv
  \gamma_B
\left(
  1
- \vec v_B \cdot \vec v_\alpha
\right) \,,
\end{equation}
where $\vec v_B \equiv \vec P_B / E_B$ is the velocity corresponding to the 
global Lorentz boost. Note that $\gamma_\alpha$ is not necessarily equal to 
$\gamma_B$. Only when the jet momentum $P_\alpha$ is perpendicular to the 
direction of the global Lorentz boost,
the two Lorentz boost factors can be the same.
The change in the dimensionless part of (\ref{eq:Jn}) can be compensated by,
\begin{equation}
  \beta \rightarrow \gamma^2_\alpha \beta \,,
\label{eq:beta-scaling}
\end{equation}
and the jet function should be redefined as,
\begin{equation}
  J^{(n)}_\beta
\rightarrow
  \gamma^{-n}_\alpha J^{(n)}_{\gamma^2_\alpha \beta} \,,
\end{equation}
to retain the original jet function value. Consequently, the clustering 
sequence remains. In this sense, the jet algorithm can be
made Lorentz invariant.

\section{Conclusion}
\label{sec:conclusion}

We reveal the direct link between the Georgi algorithms of jet clustering and the 
parton shower kinematics. The energy increases when clustering jets, due to 
conservation, while the jet mass term $P^2_\alpha / E_\alpha$ does not increase 
much, because of the fact that parton shower tends to emit soft partons, 
$z \rightarrow 0$. Our observation provides a sound support for 
this elegant class of jet clustering algorithms whose kinematic features are 
explored in this work systematically for both massless and massive partons. 
We further generalize the jet function definition to $J^{(n)}_\beta(P_\alpha)$, 
with a free jet index $n$ which is constrained within the range $1 \leq n \leq 2$. 
Its upper limit comes from the logical consistency of the jet 
algorithm, while the lower from the requirement that the cone size cannot 
be arbitrarily large in order to avoid mixing up neighbor jets. The parameter 
$\beta$ and index $n$ have the meaning of phase space boundaries and are 
constrained as $\beta > 4 / n(5-n) \geq 2/3$. 
In this generalization, the original Georgi algorithms can be recovered as special
cases, $J_\beta(P_\alpha) = J^{(1)}_\beta(P_\alpha)$. Under Lorentz boost, the
value of jet function at each step can be restored by adjusting $\beta$
and multiplying an overall factor $\gamma^{-n}$. In this sense, we claim 
that the jet algorithms can be made boost invariant.

\section{Acknowledgements}

SFG is grateful to Kaoru Hagiwara, Grisha Kirilin, and Junichi Kanzaki for 
discussions about parton shower and introduction to this field of research.
The current work is supported by Grant-in-Aid for Scientific Research
(No. 25400287) from JSPS. During the revision of this paper, the author 
received kind help from Prof. Howard Georgi to compare with his pioneering paper.

%\addcontentsline{toc}{section}{References}
%\begin{thebibliography}{99}
% Insert bibliographies here!!!
%\end{thebibliography}

\bibliographystyle{hunsrt}
\bibliography{georgi}

\begin{thebibliography}{10}

\bibitem{Weinberg77}
George~F. Sterman and Steven Weinberg.
\newblock {\it Jets from Quantum Chromodynamics}.
\newblock {\em Phys.Rev.Lett.}, 39:1436, 1977.

\bibitem{Catani93}
S.~Catani, Yuri~L. Dokshitzer, M.H. Seymour, and B.R. Webber.
\newblock {\it Longitudinally invariant $K_t$ clustering algorithms for hadron
  hadron collisions}.
\newblock {\em Nucl.Phys.}, B406:187--224, 1993.

\bibitem{Ellis93}
Stephen~D. Ellis and Davison~E. Soper.
\newblock {\it Successive combination jet algorithm for hadron collisions}.
\newblock {\em Phys.Rev.}, D48:3160--3166, 1993, [arXiv:hep-ph/9305266].

\bibitem{Dokshitzer97}
Yuri~L. Dokshitzer, G.D. Leder, S.~Moretti, and B.R. Webber.
\newblock {\it Better jet clustering algorithms}.
\newblock {\em JHEP}, 9708:001, 1997, [arXiv:hep-ph/9707323].

\bibitem{Wobisch98}
M.~Wobisch and T.~Wengler.
\newblock {\it Hadronization corrections to jet cross-sections in deep
  inelastic scattering}.
\newblock 1998, [arXiv:hep-ph/9907280].

\bibitem{WobischThesis}
M.~Wobisch.
\newblock {\it Measurement and QCD analysis of jet cross-sections in deep
  inelastic positron proton collisions at $\sqrt s$ = 300-GeV}.
\newblock 2000.

\bibitem{Cacciari08}
Matteo Cacciari, Gavin~P. Salam, and Gregory Soyez.
\newblock {\it The Anti-$k_t$ jet clustering algorithm}.
\newblock {\em JHEP}, 0804:063, 2008, [arXiv:0802.1189 [hep-ph]].

\bibitem{Catani91}
S.~Catani, Yuri~L. Dokshitzer, M.~Olsson, G.~Turnock, and B.R. Webber.
\newblock {\it New clustering algorithm for multi-jet cross-sections in $e^+
  e^-$ annihilation}.
\newblock {\em Phys.Lett.}, B269:432--438, 1991.

\bibitem{Moretti98}
Stefano Moretti, Leif Lonnblad, and Torbjorn Sjostrand.
\newblock {\it New and old jet clustering algorithms for electron - positron
  events}.
\newblock {\em JHEP}, 9808:001, 1998, [arXiv:hep-ph/9804296].

\bibitem{Blazey00}
Gerald~C. Blazey, Jay~R. Dittmann, Stephen~D. Ellis, V.~Daniel Elvira,
  K.~Frame, et~al.
\newblock {\it Run II jet physics}.
\newblock pages 47--77, 2000, [arXiv:hep-ex/0005012].

\bibitem{Ellis07}
S.D. Ellis, J.~Huston, K.~Hatakeyama, P.~Loch, and M.~Tonnesmann.
\newblock {\it Jets in hadron-hadron collisions}.
\newblock {\em Prog.Part.Nucl.Phys.}, 60:484--551, 2008, [arXiv:0712.2447
  [hep-ph]].

\bibitem{Salam09}
Gavin~P. Salam.
\newblock {\it Towards Jetography}.
\newblock {\em Eur.Phys.J.}, C67:637--686, 2010, [arXiv:0906.1833 [hep-ph]].

\bibitem{Ali10}
Ahmed Ali and Gustav Kramer.
\newblock {\it Jets and QCD: A Historical Review of the Discovery of the Quark
  and Gluon Jets and its Impact on QCD}.
\newblock {\em Eur.Phys.J.}, H36:245--326, 2011, [arXiv:1012.2288 [hep-ph]].

\bibitem{Georgi14}
Howard Georgi.
\newblock {\it A Simple Alternative to Jet-Clustering Algorithms}.
\newblock 2014, [arXiv:1408.1161 [hep-ph]].

\bibitem{Catani92}
S.~Catani, L.~Trentadue, G.~Turnock, and B.R. Webber.
\newblock {\it Resummation of large logarithms in e+ e- event shape
  distributions}.
\newblock {\em Nucl.Phys.}, B407:3--42, 1993.

\bibitem{Gieseke03}
Stefan Gieseke, P.~Stephens, and Bryan Webber.
\newblock {\it New formalism for QCD parton showers}.
\newblock {\em JHEP}, 0312:045, 2003, [arXiv:hep-ph/0310083].

\bibitem{Mueller81}
Alfred~H. Mueller.
\newblock {\it On the Multiplicity of Hadrons in QCD Jets}.
\newblock {\em Phys.Lett.}, B104:161--164, 1981.

\bibitem{Ermolaev81}
B.I. Ermolaev and Victor~S. Fadin.
\newblock {\it Log-Log Asymptotic Form of Exclusive Cross-Sections in Quantum
  Chromodynamics}.
\newblock {\em JETP Lett.}, 33:269--272, 1981.

\end{thebibliography}
\nocite{*}

\end{document}